\documentclass[twocolumn,showpacs,aps,pra,superscriptaddress]{revtex4}
\usepackage{epsfig,amsmath,graphicx,gensymb,textcomp,hyperref}

\def\w{\omega}
\def\W{\Omega}
\def\a{\alpha}
\def\({\left(}
\def\){\right)}
\def\<{\left\langle}
\def\>{\right\rangle}
\newcommand{\ud}{\,\mathrm{d}}
\newcommand{\re}[1]{\operatorname{Re}{\left\{#1\right\}}}

\newcommand{\modulus}[1]{\left| #1 \right|}

\begin{document}

\title{Quantum-limited measurements of distance fluctuations with a multimode detector}

\author{Val\'{e}rian Thiel}
\affiliation{\mbox{Laboratoire Kastler Brossel, UPMC-Sorbonne Universit\'es, CNRS, ENS-PSL Research University, Coll\`ege de France} 4 place Jussieu, 75252 Paris, France}
\affiliation{Clarendon Laboratory, Department of Physics, University of Oxford, Oxford OX1 3PU, UK}
\author{Jonathan Roslund}
\author{Pu Jian}
\author{Claude Fabre}
\author{Nicolas Treps}
\affiliation{\mbox{Laboratoire Kastler Brossel, UPMC-Sorbonne Universit\'es, CNRS, ENS-PSL Research University, Coll\`ege de France} 4 place Jussieu, 75252 Paris, France}

\date{\today}

\begin{abstract}
An experimental scheme is introduced to measure multiple parameters that are encoded in the phase quadrature of a light beam.
Using a modal description and a spectrally-resolved homodyne detection, it is shown that all of the information is collected simultaneously, such that a single measurement allows extracting the value of multiple parameters \emph{post-facto}.
With a femtosecond laser source, we apply this scheme to a measurement of the delay between two pulses with a shot-noise limited sensitivity as well as extracting the dispersion value of a dispersive medium.

\end{abstract}

\pacs{To be added}

\maketitle
\section{Introduction}
The highly synchronized optical pulse trains of mode-locked lasers have proven especially useful for precision measurements of temporal and distance displacements at the shot noise limit.  
Femtosecond pulses combine the subwavelength, interferometric precision afforded by optical coherence with a large ambiguity range offered by time-of-flight measurements of the temporal pulse train. 
These dual characteristics have found great utility in remote ranging and positioning \cite{lay2003mstar,coddington2009rapid,lee2010time,van2012many} while also being used to characterize the temporal jitter of the sources themselves for applications that include high fidelity optical time transfer \cite{giorgetta2013optical} and the creation of low phase-noise microwaves \cite{fortier2011generation}. 
The high degree of optical and RF coherence results from a coherent superposition of $\sim 10^{5} - 10^{6}$ individual frequency components. 
With this in mind, the co-propagation of a large number of coherent frequency modes offers a unique platform for studying quantum-limited, multimode measurement schemes. 

The measurement of fluctuations, whether temporal or spatial in origin, may broadly be envisioned as the precise estimation of a parameter $p$ that is encoded within the light field $E(p)$. A variation of the parameter $p$ has the effect of altering the field structure, and this change may be described as $\delta E(p) = \delta p \cdot \left. \left[ \partial E(p) / \partial p \right] \right|_{\delta p = 0}$. Hence, a fluctuating parameter $\delta p$ excites an optical mode with a structure that is given as the derivative of the static, unperturbed field with respect to the parameter of interest. 
For instance, a phase fluctuation of a continuous-wave source excites an optical mode that is in quadrature with the mean field. 
Propagation through a complex, dispersive medium encodes multiple parameters into the beam $p_{1}, p_{2}, \ldots$ (e.g., coherent phase shifts, displacements of the time-of-flight arrival, refractive index changes, etc.), and each parameter $p_{i}$ has its own optical mode $\partial E(\vec{p}) / \partial p_{i}$  \cite{jian2012real}. 
Thus, a general fluctuation $\delta \vec{p}$ induces a multimode field, in which information about the various parameters of interest is distributed among several optical modes, and each mode is a superposition of all the frequencies contained within the multispectral field.

Homodyne detection has been demonstrated to provide an optimal platform for extracting these parameters \cite{lamine2008}, which mixes the multimode field under study with a local oscillator. Mode-specificity is achieved by sculpting the spectral composition of the local oscillator into a form that matches the target mode $\partial E(\vec{p}) / \partial p_{i}$, thereby measuring the fluctuations of the parameter $p_{i}$. In this way, homodyne detection isolates a single mode and quantifies its fluctuations. 
Thus, rather than analyzing the effect of a fluctuating parameter on every frequency contained within the pulse train, it is possible to describe the perturbation with a succinct number of modes. 

The present work takes this multimodal approach as a means for describing a field subject to temporal or distance perturbations, and goes on to introduce a multidimensional homodyne system that is capable of simultaneously interrogating several modes. Quantum Cramer-Rao theory may be used to demonstrate that such a multimodal homodyne strategy is optimal (e.g., it measures the collected signal photons in the most efficient manner) \cite{pinel2012ultimate}. As such, it is possible to extract multiple parameters characterizing the light field in a single measurement with a precision limited only by the quantum nature of light. 

\section{Theoretical development}
%
    \subsection{Multimode field description} \label{sec:multimode-theory}
    We consider a linearly polarized electromagnetic field, which allows for treating the field as a scalar wave. The complex spectral field is written as $E(\w-\w_0) = A(\w-\w_0) e^{i\phi (\w) }$, where $A(\w)$ is the complex spectral amplitude of the field, $\phi(\w)$ is the spectral phase of the field, and $\w_0$ is the carrier frequency. For simplicity, the envelope is considered as a continuous distribution as opposed to a series of discrete frequencies (as are present for a pulse train). To introduce the multimode characteristic of the field, we take the example of a distance perturbation in a dispersive medium. The spectral phase is then decomposed as follows: $\phi(\w) = \phi_0(\w) + \delta\phi(\w)$, where $\phi_0(\w)$ represents the spectral phase of the unperturbed field, and $\delta\phi = \w n(\w) \delta L/c$ is the phase perturbation experienced by a variation in the propagation medium of length $\delta L$ and index $n(\w)$, where $c$ is the speed of light.

    The spectral amplitude $A(\w)$ is written as $A(\w) = \mathcal{E}_0 \a \, u_0(\w)$, where $\mathcal{E}_0$ is the field constant associated with a single photon \footnote{Under the paraxial and slowly varying envelope approximation, $\mathcal{E}_0=\sqrt{\frac{\hbar\w_0}{2nc\varepsilon_0 S T}}$ where $\varepsilon_0$ is the vacuum permittivity, $S$ is the surface area of detection and $T$ is the integration time.}
, $\a = \sqrt{N}$ (under the assumption of a classical laser field), and $u_0(\w)$ represents the normalized mode of the field \cite{grynberg2010}.
%
For a slowly varying spectral phase $\delta\phi(\w)$, it may be Taylor expanded about the carrier frequency $\w_{0}$ as 
\begin{align}\label{eq:taylorphase}
    \delta\phi(\W) = \delta\phi_0 + \W \, {\delta\phi_0}' + \frac{\W^2}{2} \, {\delta\phi_0}'' + \ldots
\end{align}
    where $\W = \w-\w_0$ is the optical frequency relative to the carrier, and the derivatives are taken relative to $\w$ and evaluated at the carrier frequency $\w_0$. For distance variations smaller than the optical wavelength (i.e., $\delta L  \ll \lambda$ ), this expansion of the phase allows the spectral field to be written as  
\begin{align}\label{eq:fieldexpanded}
    E(\W) &\simeq \mathcal{E}_0 \a
    \left[
    ( 1 + i \, \delta\phi_0 ) \,u_0(\W) \right. \nonumber\\
    &\left. + i \, {\delta\phi_0}' \, \W \, u_0(\W)
    + i \, {\delta\phi_0}'' \, \W^2 \, u_0(\W) + \ldots
    \right].
\end{align}
    Thus, in the presence of the phase perturbation, the field is no longer adequately represented by a single spectral mode $ u_{0} (\W)$, but rather with a set of spectral modes, namely $i \, u_0(\W)$, $i \, \W \, u_0(\W)$ and $i \, \W^2 \, u_0(\W)$. Consequently, a laser field experiencing a global delay is described as a superposition of the unperturbed mean field mode and higher order modes into which photons are displaced. Note that, for simplicity, we subsume the global spectral phase $\phi_0(\w)$ into the mean-field mode $u_0(\w)$, such that an observed phase originates only from the distance perturbation.
    
    After relating the phase perturbation $\delta\phi$ to a distance perturbation $\delta L$ and performing an additional Taylor expansion of the index $n(\w)$ about $\w_{0}$, Eq.\eqref{eq:fieldexpanded} may be rewritten as
\begin{align}\label{eq:fieldexpanded2}
    E(\W) \simeq \mathcal{E}_0 \a
    \left[
     u_0(\W) + \w_0 \cdot t_\varphi \cdot u_\varphi(\W)
    + \Delta\w \cdot t_g \cdot u_g(\W)
    \right],
\end{align}
    where $\Delta\w$ is the square root of the second moment of the spectral intensity (i.e., $\Delta \w ^{2} = \int \modulus{u_{0}(\W)}^{2} \ud \W$) and the normalized modes $u_\varphi(\W) = i \, u_0(\W)$ and $u_g(\W) = i \, \frac{\W}{\Delta\w} u_0(\W)$ have been introduced \footnote{The modes are normalized according to $\int \ud \w \; u_i^\ast(\w) u_j(\w) = \delta_{ij}$}. The variables $t_\varphi = n(\w_0) \delta L / c$ and $t_g = \big(n(\w_0) + \w_0 n'(\w_0) \big) \delta L / c$ describe a temporal slippage of the optical carrier and the pulse envelope, respectively. 
        
        This result, of course, corresponds to the fact that a global delay of the field is achieved by delaying both the optical carrier $t_\varphi $ and the temporal envelope $t_g$, and these individual delays are not necessarily equal in a dispersive medium. The modal concept, however, allows for independently decomposing this global temporal delay in terms of its isolated effect on the optical carrier and envelope function, which is achieved by analyzing the number of photons displaced into the carrier $u_\varphi(\W)$ and envelope $u_g(\W)$ spectral modes.
    \subsection{Projective measurements}

Under the general conclusion that a perturbation transforms a single mode field into a multimode one, the signal field under study $E_s(\W)$ may be more generally expanded in a basis of spectral modes $\left\{ v_n \right\}$:
\begin{align}\label{eq:fieldexpandedgeneral}
    E_s(\W) & =\mathcal{E}_0 \, \alpha_s \, \( u_0 (\W) + \sum_n p_n \, K_n \, v_n(\W) \) \nonumber\\
    & \equiv \mathcal{E}_0 \, \alpha_s \, u_s(\W) , 
\end{align}
where $u_{0}(\W)$ once again represents the mode of the unperturbed field and $u_s(\W)$ labels the mode of the entire field. The normalized modes $v_n$ are defined by $v_n = \frac{1}{K_n} \frac{\partial u_s}{\partial p_n}$ \footnote{Note that such a definition of the modes does not necessarily imply that they form a complete, orthonormal basis; in the general case, these modes are not orthogonal.}, in which the normalization factor $K_n$ is given by $K_{n} =  \left( \int \ud\W \, \partial u_s^\ast(\W) / \partial p_{n} \cdot \partial u_s(\W) / \partial p_{n} \right)^{1/2}$, and $p_n$ is the parameter of interest that is to be extracted \cite{jian2012real}. 
Thus, a perturbation of the field variable $p_{n}$ displaces photons into a sideband with a modal structure given by the derivative of the field with respect to $p_{n}$. This expansion is readily seen to correspond to a generalization of Eq.~\ref{eq:fieldexpanded2}. 

    Homodyne detection provides a mode-sensitive detection methodology that is capable of retrieving the information encoded in a single parameter $p_{n}$. In particular, the multimode signal field $E_s(\W)$ is mixed with a stronger field, termed the local oscillator (LO) in the mode $u_{LO}(\W)$, on a $50:50$ beamsplitter. The light exiting each of the two output ports of the beamsplitter is then detected with individual photodiodes \cite{bachor2004guide}. The homodyne signal is obtained by taking the difference of the photocurrents from these two diodes and reads as $S_\textrm{HD} = R \cdot \int \( E_\textrm{LO} E_\textrm{s}^\ast + \textrm{c.c} \) \ud \vec{r} \, \ud \W$, where $R$ represents the responsivity of the photodetection scheme \cite{bachor2004guide}. 
    The diode responsivity $R$ is subsequently taken to be unity as it does not affect the resultant signal-to-noise (assuming shot noise limited detection). 

By using the field expression of Eq.~\ref{eq:fieldexpandedgeneral} and assuming that the signal field is a coherent state (i.e., $\alpha = \sqrt{N}$), the homodyne signal is written as
\begin{align}\label{eq:HDsig1}
    S_\textrm{HD} = 2 \, \mathcal{E}_0^{2} \, \sqrt{N_\textrm{LO} \, N} \, \re{ \eta \, e^{i\phi_\emph{rel}} }
\end{align}
%
    where $\eta =  \int \ud\W \, u_s^\ast(\W) \cdot u_\textrm{LO}(\W)$ is the overlap between the signal and LO spectral modes, and $\phi_{\emph{rel}}$ is the relative, spectrally-independent phase between the two fields (any relative, spectrally-dependent phase between the fields is contained in $\eta$, which represents the temporal / spectral overlap between the two fields). Hence, homodyne detection may be viewed as a projective measurement scheme. In this manner, an individual parameter $p_{n}$ is measured by sculpting the frequency structure of the LO field to match the corresponding signal mode $v_n$.

    In what follows, the spatial overlap of the two fields is taken to be unity (i.e., the two fields are in the same spatial mode, which is experimentally ensured by filtering the output ports of the beamsplitter with a single mode fiber) as it does not affect the qualitative understanding of the spectral projection technique. However, a reduction in the spatial overlap implies that fewer photons are detected in the signal mode of interest. This effect is taken into account when determining the theoretical values for this paper.

If the relative phase is set to be $\phi_{rel} = 0$ and the LO mode is also shaped to perfectly overlap with a signal mode $v_{n}$ of the signal field (i.e., $u_{\textrm{LO}} = v_{n}$), the homodyne signal is given by 
\begin{align}\label{eq:HDsig2}
    S_\textrm{HD} = 2 \, \mathcal{E}_0^{2} \,\sqrt{N_\textrm{LO} \, N} \, p_n K_n.
\end{align}

    Note that it has also been implicitly assumed that the $v_n$ are all orthogonal and form a basis, which is not necessarily the case since these result from derivatives of the field. In such a case, the LO may overlap with multiple modes. In general, it is possible to orthogonalize the modes $v_{n}$ if a parameter of interest is contaminated by nonzero overlap with a mode corresponding to a different parameter \cite{jian2012real}.
    
    With the assumption that $N_{\textrm{LO}} \gg N$, the noise variance in the measurement is given by $\Delta S_\textrm{HD} = \mathcal{E}_0^{2} \, \sqrt{N_\textrm{LO}} \, \sigma_n$, where $\sigma_n$ is the noise variance of the signal mode $n$ relative to the quantum limit (i.e., $\sigma_{n} = 1$ for a quantum-noise limited beam and $\sigma_{n} > 1$ when technical noise is present) \cite{grynberg2010}. Hence, the signal-to-noise (SNR) ratio $\Sigma_n$ is given by
\begin{align}\label{eq:HDsigSNR}
    \Sigma_n = \frac{S_\textrm{HD}}{\Delta S_\textrm{HD}}= 2 \, p_n \, \frac{\sqrt{N} \, K_n}{\sigma_n},
\end{align}
    which increases as the square root of the optical power in the signal field. The case where noise in the $n^\textrm{th}$ mode is quantum-limited (i.e., $\sigma_{n} = 1$) results in the highest SNR obtainable using classical resources and is termed the Standard Quantum Limit (SQL).
    \subsection{Application to ranging} \label{sec:theory-ranging}

    We are now in a position to examine the means by which projective measurements are capable of extracting mode-specific information for the measurement of distance fluctuations. The minimal detectable fluctuation of a parameter $p_{n}$ is given by a SNR of $\Sigma_{n} = 1$, which corresponds to $p_{n,\textrm{min}} = 1 / 2 \sqrt{N} \, K_{n}$.    
    Recalling the expansion of Eq.~\eqref{eq:fieldexpanded}, a projection of the perturbed signal field $E_{s}(\W)$ onto the mode $u_\varphi (\W)$ is capable of detecting a minimal optical carrier slippage $t_\varphi$ of $\( t_\varphi \)_\textrm{min} = \frac{1}{2\sqrt{N}\w_0}$. Conversely, a projection onto the envelope-sensitive mode $u_g(\W)$ measures a group delay $t_g$ with a sensitivity of $\( t_g \)_\textrm{min} = \frac{1}{2\sqrt{N}\Delta\w}$.
    
    As expected, the temporal position of the optical carrier may be pinpointed with higher precision than that of the envelope due to the fact that the optical frequency $\w_{0}$ is larger than the pulse bandwidth $\Delta \w$. Thus, projection of the signal field onto $u_\varphi (\W)$ and $u_g(\W)$ is equivalent to a coherent interferometric measurement of an optical fringe and an incoherent time-of-flight measurement of an envelope function, respectively. The former has increased precision but an inability to identify the absolute optical cycle while the later has a larger dynamic range at the expense of precision.

In the case of a non-dispersive medium (i.e., $n'(\w_0) = 0$), a change in the optical path leads to a commensurate retardation of the optical carrier and envelope, i.e., $t_\varphi = t_{g}$. Hence, a projective measurement of $u_\varphi (\W)$ and $u_g(\W)$ reveals identical information, albeit with different sensitivities. 
Rather that separately considering the delays of just the optical carrier $t_\varphi$ or the temporal envelope $t_{g}$, it is possible to define an additional mode $v_{ t}$ that is associated with   
the global delay $t_\varphi = t_{g}$ which carries information about the overall distance variation $\delta L$. This detection mode $v_{\delta L}$ is obtained by differentiating the total signal field of Eq.~\eqref{eq:fieldexpanded} with respect to $\delta L$. With the use of Eq.~\eqref{eq:fieldexpanded2}, it is straightforward to see that $v_{\delta L} \propto \w_0 \, n(\w_{0}) \, \cdot u_\varphi(\W) + \Delta\w \, n_{g} \, \cdot u_g(\W)$ where the group index $n_{g}$ is given as $n_{g} = n(\w_0) + \w_0 \, n'(\w_0)$. Thus, the detection mode $v_{\delta L}$ corresponds to a weighted superposition of the carrier $u_\varphi(\W)$ and envelope $u_{g} (\W)$ modes. Following normalization, the distance detection mode is written as:
\begin{align}\label{eq:detmodeL}
    v_{\delta L}(\W) = \frac{1}{ \sqrt{ \(\frac{\w_0}{\textrm{v}_\varphi}\)^2 + \(\frac{\Delta\w}{\textrm{v}_g}\)^2 } } \( \frac{\w_0}{\textrm{v}_\varphi} \cdot u_\varphi(\W) + \frac{\Delta\w}{\textrm{v}_g} \cdot u_g(\W) \),
\end{align}
    where $\textrm{v}_\varphi = c/n(\w_0)$ and $\textrm{v}_g = c/n_{g}$ are respectively the phase and group velocities of light in the propagation medium, which can be considered as equal in the case of air.
    A projection of the signal field onto this mode then measures the global displacement $\delta L$ with a sensitivity of 
\begin{align}\label{eq:sensL}
    \big(\delta L\big)_\emph{SQL} = \frac{1}{ 2\sqrt{N} }\frac{1}{ \sqrt{ \(\frac{\w_0}{\textrm{v}_\varphi}\)^2 + \(\frac{\Delta\w}{\textrm{v}_g}\)^2 } }.
\end{align}

Importantly, the sensitivity of this timing mode exceeds that of both the carrier $u_\varphi(\W)$ and envelope $u_{g} (\W)$ modes, and the enhancement of precision relative to that obtained from the carrier mode is $\sqrt{1 + \left(\Delta \w / \w_{0} \right)^{2} }$. Although this enhancement is small for modest bandwidths (i.e., $\Delta \w / \w_{0} \ll 1$), it nevertheless illustrates the fact that the best sensitivity for measuring longitudinal displacements is obtained by combining an interferometric and a time-of-flight measurement. 
\section{Methodology}
%
    \subsection{Experimental layout}

The present work employs a Mach-Zehnder interferometer for the creation and detection of phase perturbations $\delta\phi(\W)$. In particular, the output of a commercial Titanium-Sapphire femtosecond oscillator delivering 22 fs pulses at 795 nm is divided with a $90:10$ beamsplitter. The $90\%$ port of the beamsplitter constitutes the strong LO field while the weaker $10\%$ port is the signal field $E_{w}(\W)$.  

The signal field is displaced longitudinally to create the phase perturbation, which is accomplished by 
reflecting the beam from a zero-degree mirror that is mounted upon a piezo actuator. The piezo element is modulated to create a high frequency ($\sim 2 \textrm{MHz}$) phase perturbation. The absence of any technical noise at this frequency facilitates shot-noise limited detection.
A pulse shaper is present in the LO arm of the interferometer in order to compensate for any differential spectral phase between the LO and signal arms. This dispersion compensation enhances the mean temporal / spectral overlap between the two fields at the second beamsplitter.

\begin{figure}[tbp]
    \centering
    \includegraphics[width=7.5cm]{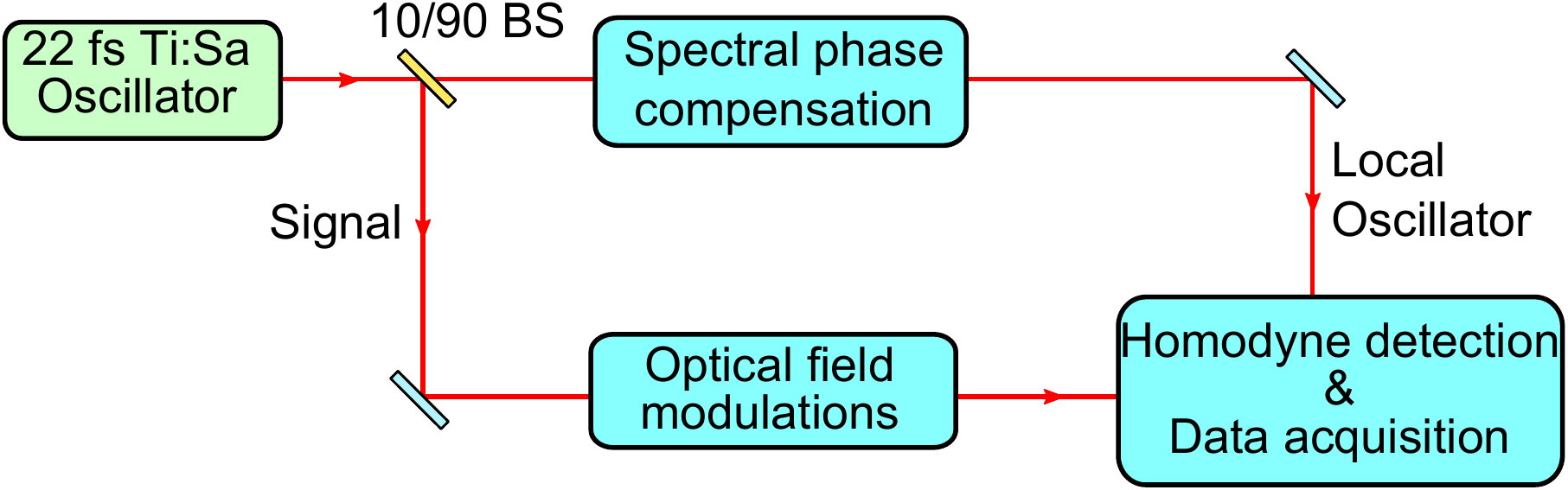}
    \caption{
    General experimental layout. A femtosecond oscillator is split on a 10/90 beamsplitter (BS) into the signal field and the local oscillator field. A phase perturbation is introduced on the signal field; on the local oscillator, a pulse shaper is utilized to compensate the spectral phase between the two paths.
    }
    \label{fig:exp-layout}
\end{figure}

The two fields and then recombined on a $50 : 50$ beamsplitter, in which each output port is independently detected with a photodiode. In order to ensure perfect spatial overlap between the LO and signal fields, the combined fields at the output of the beamsplitter are filtered with single mode fibers prior to photodetection. 
Measurement of phase variations requires that the relative phase between the two interferometer arms be set to $\phi_{rel} = \pi / 2$ (there is a relative phase of $\pi / 2$ between the unperturbed field $u_0(\W)$ and the carrier and group modes). 
This phase offset is achieved by loosely locking (i.e., a lock bandwidth of $\lesssim 100 \textrm{Hz}$) the difference of the two photodiode signals with a commercial servo controller, which acts on an additional slowly-moving piezo actuator in the interferometer. The general experimental scheme is depicted in Fig.~\ref{fig:exp-layout}.
%
    \subsection{Detection scheme}
\begin{figure}[t]
	\centering
	\includegraphics[width=7.5cm]{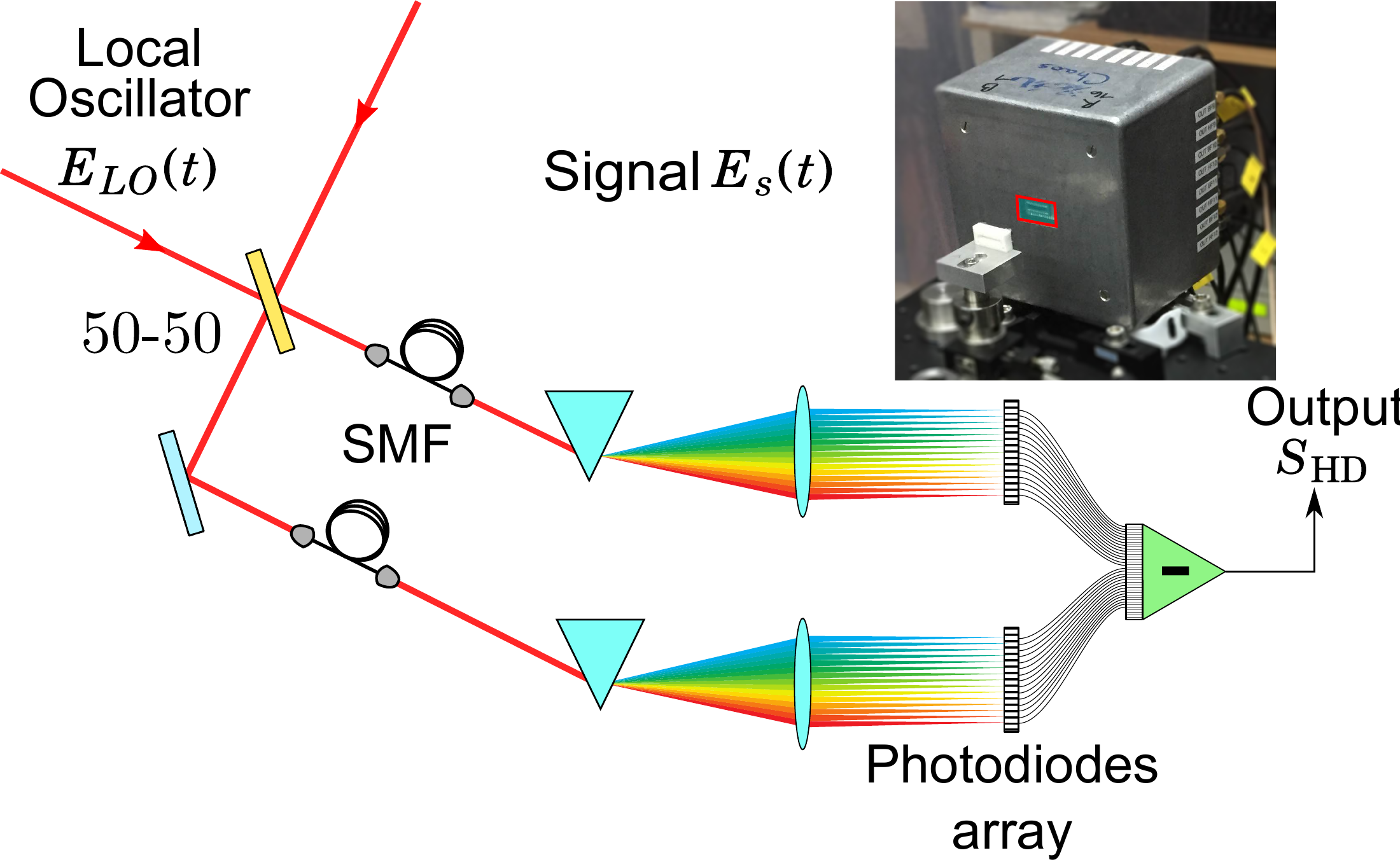}
	\caption{
		Spectrally-resolved homodyne detection scheme at the interferometer's output. SMF: single-mode fiber for spatial filtering. Inset: one detector with the photodiode array highlighted. 
	}
	\label{fig:mhdscheme}
\end{figure}

%
Since the various modes underlying the perturbed field of Eq.~\ref{eq:fieldexpanded2} are differentiated based upon their spectral composition, a frequency-resolved detection system is necessary in order to isolate their individual contributions to the overall homodyne signal. Toward that end, each output port of the $50:50$ beamsplitter is spectrally resolved with a diffraction grating and imaged onto an eight element photodiode array as seen in Fig.~\ref{fig:mhdscheme}.
The photocurrents are split into low- (DC) and high- (HF) frequency components and independently amplified. 
The HF signal is demodulated with an RF tone at the known modulation frequency and subsequently digitized using a computer-controlled acquisition card. 

Specific spectral modes may be reconstructed by taking a weighted superposition of the eight independently detected HF signals. By digitally applying a variable gain $g_{i}$ to pixel $i$, the homodyne signal of Eq.~\eqref{eq:HDsig1} may be written as
\begin{align}\label{eq:HDsigMP}
    S_\textrm{HD} \propto  \sum_{n} p_{n} \sum_{i} g_i \int_{\Delta\W_i} \ v_{n} u_{LO} \, \ud\W_i,
\end{align}
%
%
	where the overlap integral is computed over the spectral domain $\Delta\W_i$ incident on a pixel element. This is equivalent to equation \eqref{eq:HDsig2} in the limit that $\Delta\W_i$ tends to a single frequency and generally remains a good approximation in the present situation of eight distinct spectral zones.
	
    The multipixel homodyne signal from equation \eqref{eq:HDsigMP} shows that one can optimise the retrieval of a given $p_n$ for a well-chosen gain function $g$. This gain function is constructed from knowledge of the mean-field, which is obtained from the DC output from the multipixel detectors. For instance, the time-of-flight mode $v_1$ is defined as the derivative of the mean-field mode $v_0$.
%
\section{Results} 
%
\subsection{Illustration of the multimode structure} \label{sec:multimode}
%

\begin{figure}[t!]
	\centering
	\includegraphics[width=.95\linewidth]{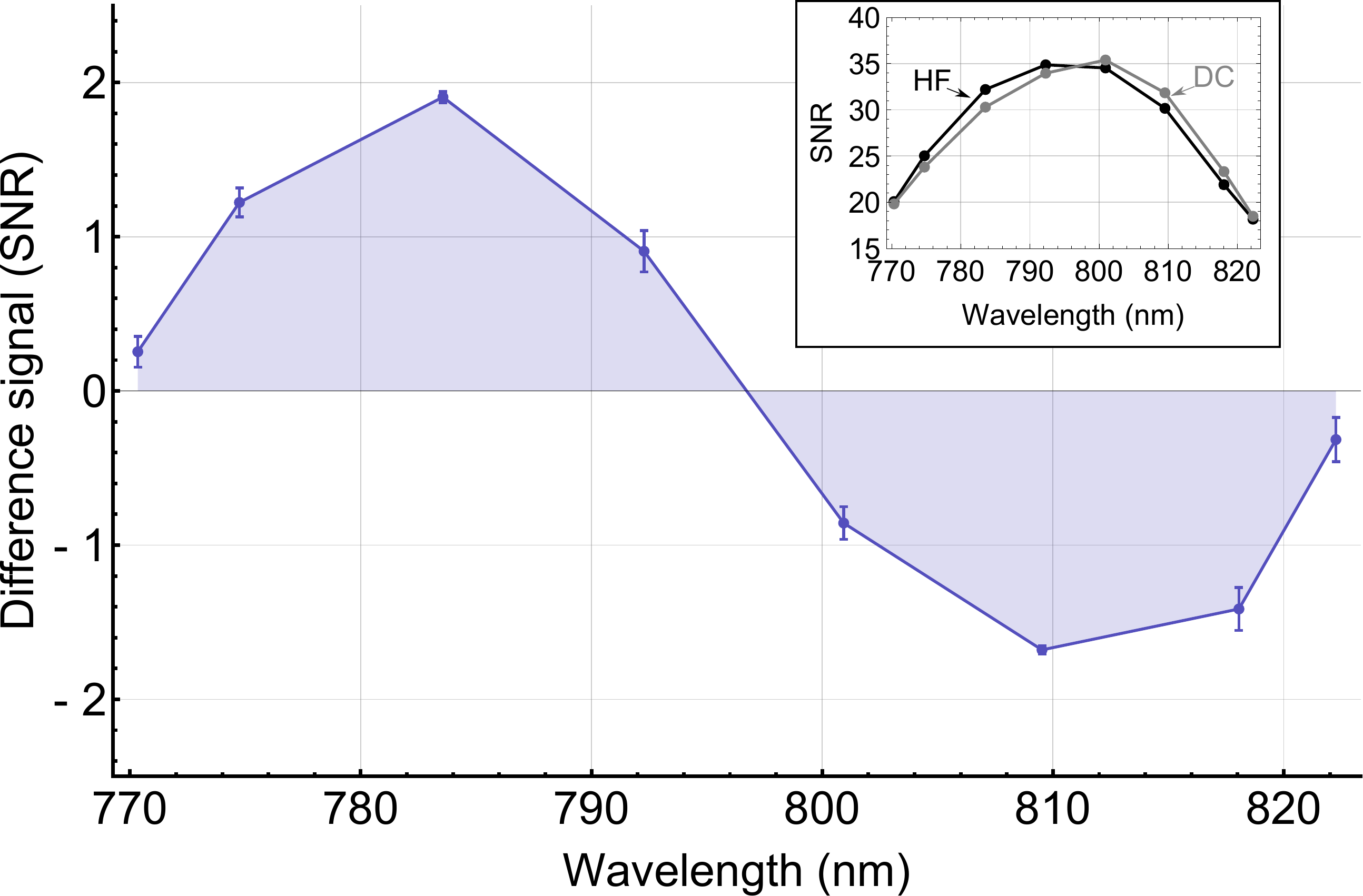}
	\caption{
		Signal-to-noise $D$ obtained by subtracting the mean-field mode from the homodyne signal. The spectral structure strongly resembles that of a time-of-flight mode, with a SNR two orders of magnitude smaller than the homodyne signal which is dominated by the mean-field mode. Error bar: standard deviation. Inset: individual terms from the difference spectrum, namely the demodulated HF signal $\Sigma$ and the DC signal $G\cdot \sqrt{P}$.
	}
	\label{fig:diffsignal}
\end{figure}

As previously described and indicated by Eq.~\ref{eq:fieldexpanded2}, the perturbation transforms the initially single mode field into a multimode structure. In order to illustrate this multimode structure, the SNR of the demodulated HF signal is shown for each pixel in the inset of Fig.~\ref{fig:diffsignal}. To a first approximation, the strength of the phase modulation measured with the homodyne approach should be proportional to the square root of the number of signal photons falling upon the detector (as in Eq.~\ref{eq:HDsigSNR}). Indeed, the spectral distribution of the HF SNR appears quite similar to the DC signal of the signal field (seen as the inset in Fig.~\ref{fig:diffsignal}).

However, according to Eq.~\ref{eq:fieldexpanded2}, the fractional contribution of the envelope mode $u_{g} (\W)$ relative to the carrier mode $u_\varphi(\W)$ is given by the ratio of the bandwidth to the carrier frequency, which is  $\Delta \w / \w_{0} \lesssim 2 \%$ in the present configuration. Thus, while a multimode structure should be present in the recovered phase perturbation, it is largely obscured due to the unbalanced contribution from the carrier and group modes. In order to verify the predicted structure, it is helpful to remove the contribution of the dominant carrier mode $u_\varphi(\W)$ from the measured signal. 

As the carrier mode is simply a phase-shifted version of the unperturbed static field, i.e., $u_\varphi(\W) = i \, u_0(\W)$, the two have an identical spectral form. Furthermore, $u_{0}(\W)$ is independently assessed with the DC signal of the signal field $P_{s}(\W)$, i.e., $u_{0}(\W) \propto \sqrt{P_{s}(\W)}$. 
If the overall SNR is rewritten in terms of its two components, i.e., $\Sigma = \Sigma_{\varphi} \cdot u_\varphi(\W) + \Sigma_{g} \cdot u_{g} (\W)$, it is possible to relate the leading carrier term to the DC signal with the relation $\Sigma_{\varphi} \cdot u_\varphi(\W) = G \cdot \sqrt{P(\W)}$ where $G$ is an undetermined gain factor. 
The overall SNR is then multiplied by the gain weighted DC term and integrated to yield: $G \int d \W \, \Sigma(\W) \cdot \sqrt{P(\W)} = \Sigma_{\phi} ^{2}$ where the identities $\int d\W \, u_{\phi}(\W) \, u_{0}(\W) = 1$ and $\int d\W \, u_{g}(\W) \,  u_{0}(\W) = 0$ have been exploited \footnote{By definition, $\int d\W \, u_{0}(\W) \, u_{g} (\W) = \int d \W \, (\W / \Delta \w) \, u_{0}^{2} = 0$.}. Using the original definition of $G$, this result establishes the equality: $G \int d \W \, \Sigma(\W) \cdot \sqrt{P(\W)} = G^{2} \int d\W \, P(\W) $. Consequently, the gain factor $G$ may be computed from the independently collected HF and DC data to be $G = \int d\W \, \Sigma(\W) \, \sqrt{P(\W)} / \int d\W \, P(\W)$. 

Finally, the difference spectrum $D(\W) = \Sigma - G \cdot \sqrt{P(\W)} = \Sigma _{g}  \cdot u_{g} (\W)$ permits examining the measured SNR without a contribution from the carrier, and the remaining structure is shown in Fig.~\ref{fig:diffsignal}. If the phase perturbation did not displace photons into the group mode $u_{g}(\W)$, the difference spectrum $D(\W)$ would be zero. However, the observed spectrum is clearly structured and resembles that of the time-of-flight mode $u_{g}(\W) = (\W / \Delta \w) \cdot u_{0} (\W)$. Thus, the phase displacement activates at least two spectral modes. Moreover, the observed SNR for $D(\W)$ indicates that this mode represents less than $2\%$ of the full signal, in agreement with the contribution predicted from Eq.~\ref{eq:fieldexpanded2}.

In the following section, we employ a more quantitative approach to project the homodyne signal onto a user-defined spectral mode.
    
    \subsection{Range-finding}
%
\begin{figure}[t!]
	\centering
	\includegraphics[width=.95\linewidth]{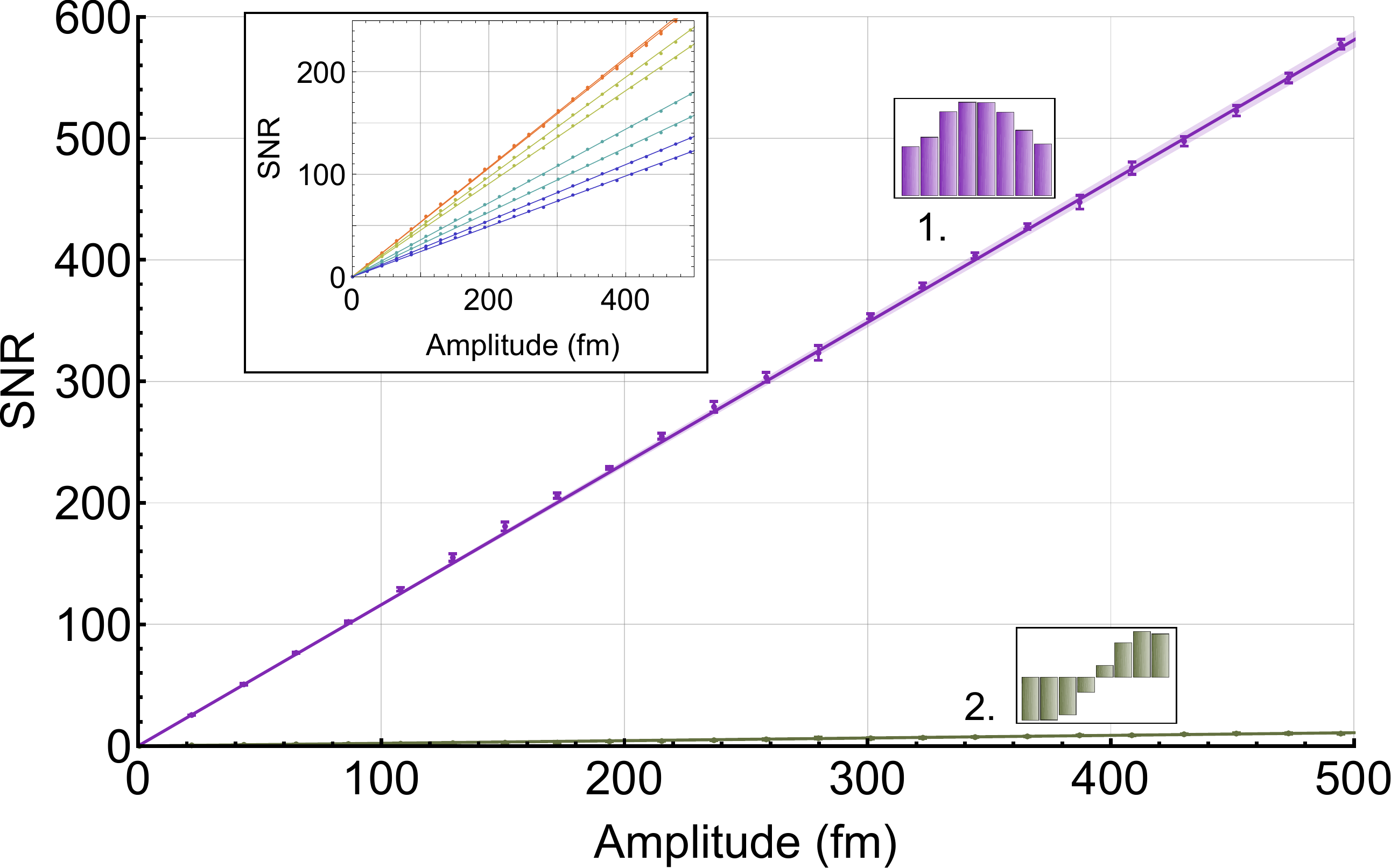}
	\caption{
		Signal-to-noise ratio for the detection of a ramped distance modulation in one arm of the interferometer. Inset: SNR for individual pixels of the detector (colour scheme represents the received optical power). Applying the detection modes to each signal and summing the result yields the traces 1:phase 2:time-of-flight. The error bars depict one standard deviation.
	}
	\label{fig:plotratio}
\end{figure}
    We now investigate the use of projective measurements for the measurement of timing / distance fluctuations between the two arms of the interferometer. As before, these fluctuations are induced with a sinusoidal modulation of the piezo actuator in order to ensure that the only contributing noise source is quantum in origin. In particular, the modulation amplitude is linearly ramped as a means for measuring the minimal displacement for various spectral modes. The mean homodyne signal for each of the eight pixels of the photodiode array is then collected as a function of the modulation amplitude. Likewise, an analysis of the homodyne variance for each pixel reveals the shot noise level, which permits determining the SNR in each spectral band.

    In this manner, the measured SNR for each pixel as a function of the modulation amplitude is shown in the inset of Fig.~\ref{fig:plotratio}. The SNR follows the square root of the number of signal photons collected by each detector element, which explains the fact that the SNR for the side pixels is lower than that recorded in the center pixels (blue to red colour scheme on the plot inset).
For reference, the modulation of the piezo actuator has been calibrated using the method outlined in \cite{thiel2016} and exhibits a displacement of $\sim 0.55 \textrm{\AA} / \textrm{Volt}$. 
    
    By taking linear combinations of these individual SNRs, it is possible to determine the sensitivity for various spectral modes with a particular interest on the carrier $u_\varphi(\W)$ and group $u_{g} (\W)$ modes. As previously explained, projection onto the carrier phase mode is equivalent to an interferometric phase estimation while projection onto the group mode is tantamount to a time-of-flight measurement. Thus, it is expected that the phase mode projection should exhibit a detection sensitivity greater than that afforded by the TOF by a factor of $\sim \w_{0} / \Delta \w$. 
    
As explained Sec.~\ref{sec:multimode-theory}, the carrier mode possesses an identical spectral structure to that of the unperturbed mean field $u_0(\W)$, which, in turn, is assessed with an independent measurement of the DC signal for the signal field $P_{s}(\W)$, i.e., $u_{0}(\W) \propto \sqrt{P_{s}(\W)}$. With the mean field spectrum in hand, the carrier frequency is computed as $\w_{0} =  \int \ud \w \, \w \cdot u_{0} (\w-\w_0)^{2}$. The group mode $u_{g} (\W)$ is then formed as $u_{g}(\W) \propto \left(\w - \w_{0} \right) \cdot u_{0}(\w-\w_0)$, which is orthogonal to the carrier mode $u_\varphi(\W)$ by construction. After normalization, they are utilized to project the carrier and envelope longitudinal displacements from the data seen in the inset of Fig.~\ref{fig:plotratio}.

    Projection onto the phase mode (blue trace of Fig.~\ref{fig:plotratio}) results in a strong SNR since the contribution from the center pixels dominate the sum. Conversely, a projection from the time-of-flight mode yields a smaller signal as seen in the red trace of Fig. ~\ref{fig:plotratio}. Each of these traces is fit to a linear form, and the displacement at which $\Sigma = 1$ corresponds to the smallest detectable displacement. 
The detection limit for the carrier mode (blue trace) is $5.6 \pm 0.1$ fm/$\sqrt{\textrm{Hz}}$
while that of the time-of-flight mode (red trace) is $300 \pm 20$ fm/$\sqrt{\textrm{Hz}}$ \footnote{The HF signals are demodulated with a 15kHz bandwidth surrounding the modulation frequency of 2MHz, and this resolution bandwidth is used for normalizing the detection sensitivity. A factor of $\pi/2$ is needed to correct the resolution bandwidth in order to define the signals in a box filter instead of that defined by a first-order filter.} 
    
    Expressed in the temporal domain, these two sensitivities are $\sim 19 \, \textrm{ys} / \sqrt{\textrm{Hz}}$ and $\sim 1000 \, \textrm{ys} / \sqrt{\textrm{Hz}}$ for the carrier and TOF modes, respectively. These values are on par with what has been observed in previous studies examining timing jitter with techniques formally equivalent to the present TOF measurement: $\sim 550 \,\textrm{ys} / \sqrt{\textrm{Hz}}$ in \cite{hou2015timing} and $\sim 1730 \, \textrm{ys} / \sqrt{\textrm{Hz}}$ in \cite{kim2011sub}.
    
    These experimentally measured limits may be compared with what is expected from the theoretical limits derived in Sec.~\ref{sec:theory-ranging}.
    For a detected optical power of $\approx 210$ \textmu $\textrm{W}$ in the signal beam \footnote{
    	Taking into account the $70\%$ quantum efficiency of the photodetectors and the $90\%$ fringe visibility. The total power is $\approx 210$ \textmu W, and the integration time is $9.5$ \textmu s, which is deduced from the resolution bandwidth of $15$ kHz obtained by the low-pass filter at the output of the demodulation stage.
    	}, the predicted sensitivity for the carrier mode is $L_{\varphi,\textrm{min}} = c / 2 \, \w_{0} \, \sqrt{N} = 5.8 \pm 0.1 \, \textrm{fm} / \sqrt{\textrm{Hz}}$ while that of the TOF mode is $L_{g,\textrm{min}} = c / 2 \, \Delta \w \, \sqrt{N} = 307 \pm 5 \, \textrm{fm} / \sqrt{\textrm{Hz}}$, which are in good agreement with the experimental values. 
The results are similar to that which was observed in \cite{thiel2016} with non-spectrally-resolved detection.

It is also possible to examine the difference in measurement precisions for the carrier and TOF modes, which is seen to be $L_{\varphi,\textrm{min}} / L_{g,\textrm{min}} = 53 \pm 3$ and is in good agreement with the expected ratio of $\w_0 / \Delta\w \approx 53.2$.

\subsection{Detection mode}
%
\begin{figure}[t!]
	\centering
	\includegraphics[width=.95\linewidth]{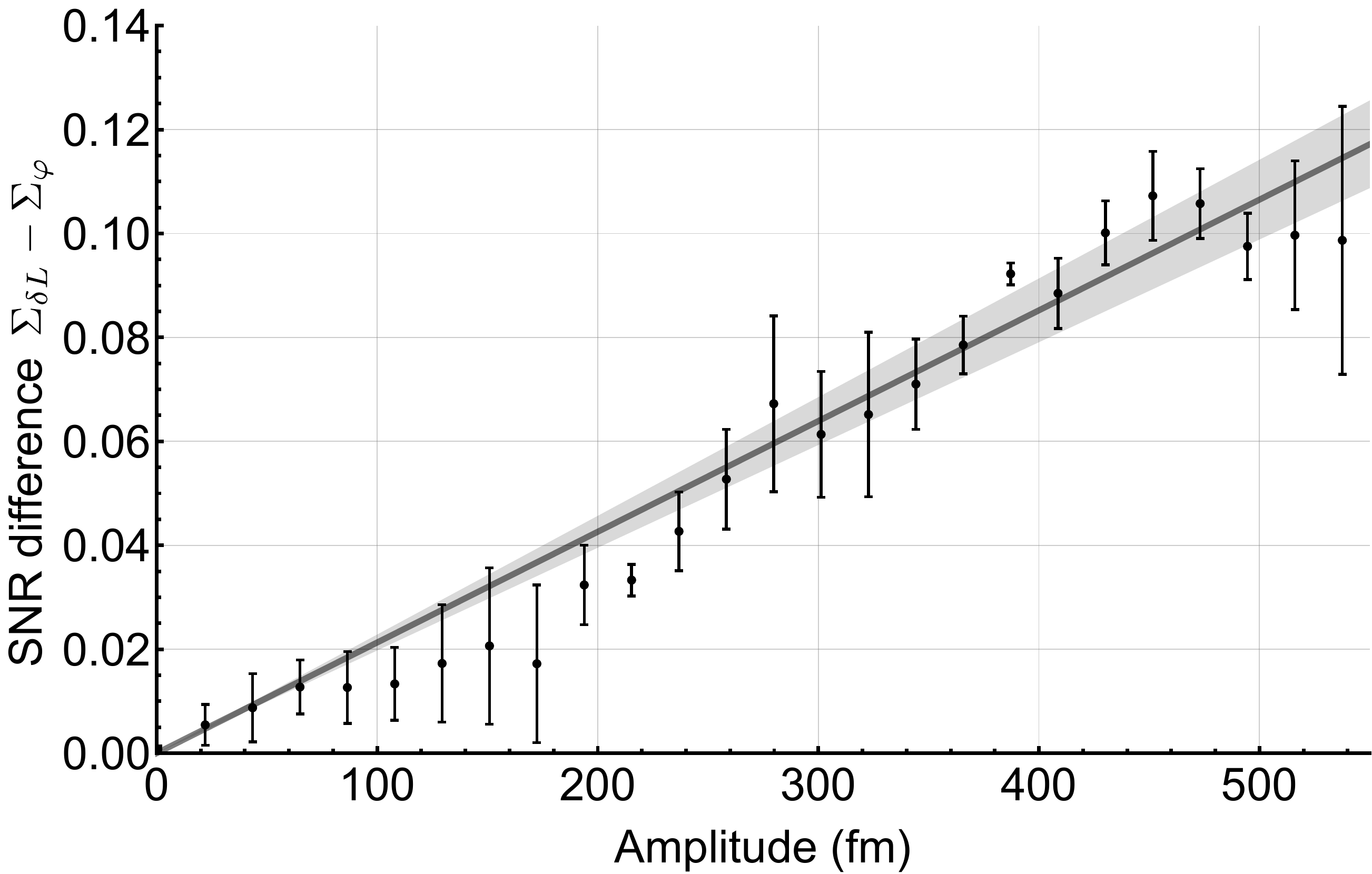}
	\caption{
		Enhancement in signal-to-noise from a projection on the detection mode versus a projection on the phase mode, $\Sigma_{\delta L} - \Sigma_\varphi$. The error bars show one standard deviation.
	}
	\label{fig:enhancement}
\end{figure}
	Sec.~\ref{sec:multimode-theory} illustrated that timing / distance fluctuations may be retrieved by projecting the perturbed multimode field onto a spectral mode that measures either a displacement of the optical carrier or the temporal envelope. Both of the examined modes provide the same information (i.e., the longitudinal displacement) but with distinct precisions. 
	
However, as explained in Sec.~\ref{sec:theory-ranging}, the ultimate sensitivity limit for the determination of timing / distance fluctuations is obtained by combining interferometric and time-of-flight measurements, which is achieved by projecting the perturbed field onto a linear combination of the carrier and group spectral modes. Specifically, this linear combination is given by Eq.~\ref{eq:detmodeL}, and the associated sensitivity is predicted to be $\delta L_\textrm{SQL} = c/2\sqrt{N} \cdot \sqrt{\w_0^2 + \Delta\w^2}$. As detailed in Sec.~\ref{sec:theory-ranging}, the sensitivity provided by the carrier mode is $\delta L_\varphi = c/2\sqrt{N} \w_0$. Thus, an amalgamation of the information provided by the carrier and TOF modes is predicted to enhance the measurement sensitivity by a factor of $\left( \delta L_\varphi - \delta L_\textrm{SQL} \right) / \delta L_\textrm{SQL} = \sqrt{1 + (\Delta \w / \w_{0} )^2} - 1 \simeq \Delta \w^{2} / 2\, \w_{0}^2 \simeq 1.8 \cdot 10^{-4}$.
	
	In order to explore this possibility, we project the spectrally-resolved SNRs seen in the inset of Fig.~\ref{fig:plotratio} onto both the carrier mode as well as the detection mode specified by Eq.~\ref{eq:detmodeL}. The difference in the associated SNRs between these two measurements $\Sigma_{\delta L} - \Sigma_\varphi$ is shown in Fig.~\ref{fig:enhancement}.
	
The detection mode exhibits a modest, although statistically meaningful, increase in sensitivity relative to that observed for the carrier mode. The experimental efficiency gain afforded by measuring with this hybrid spectral mode is found to be $(\Sigma_{\delta L}/\Sigma_\varphi) - 1 = \left( 1.8 \pm 0.2 \right) \cdot 10^{-4} $, which is in excellent agreement with the theoretical value expected from the carrier-bandwidth ratio of $\w_0 / \Delta\w \approx 53.2$.

Importantly, this increase in measurement precision is achievable due to the fact that the perturbed field is actually multimode in origin as demonstrated in Fig.~\ref{fig:diffsignal}. As a result, a disruption of the mean field structure can not be adequately represented by a displacement of the optical carrier alone (although this is the dominant contributor to the signal). By combining the interferometric and TOF approaches, it is possible to   measure timing / distance displacements with a sensitivity not obtainable with either approach in isolation. 
%
    \subsection{Index dispersion}
%
%
\begin{figure}[t!]
	\centering
	\includegraphics[width=\linewidth]{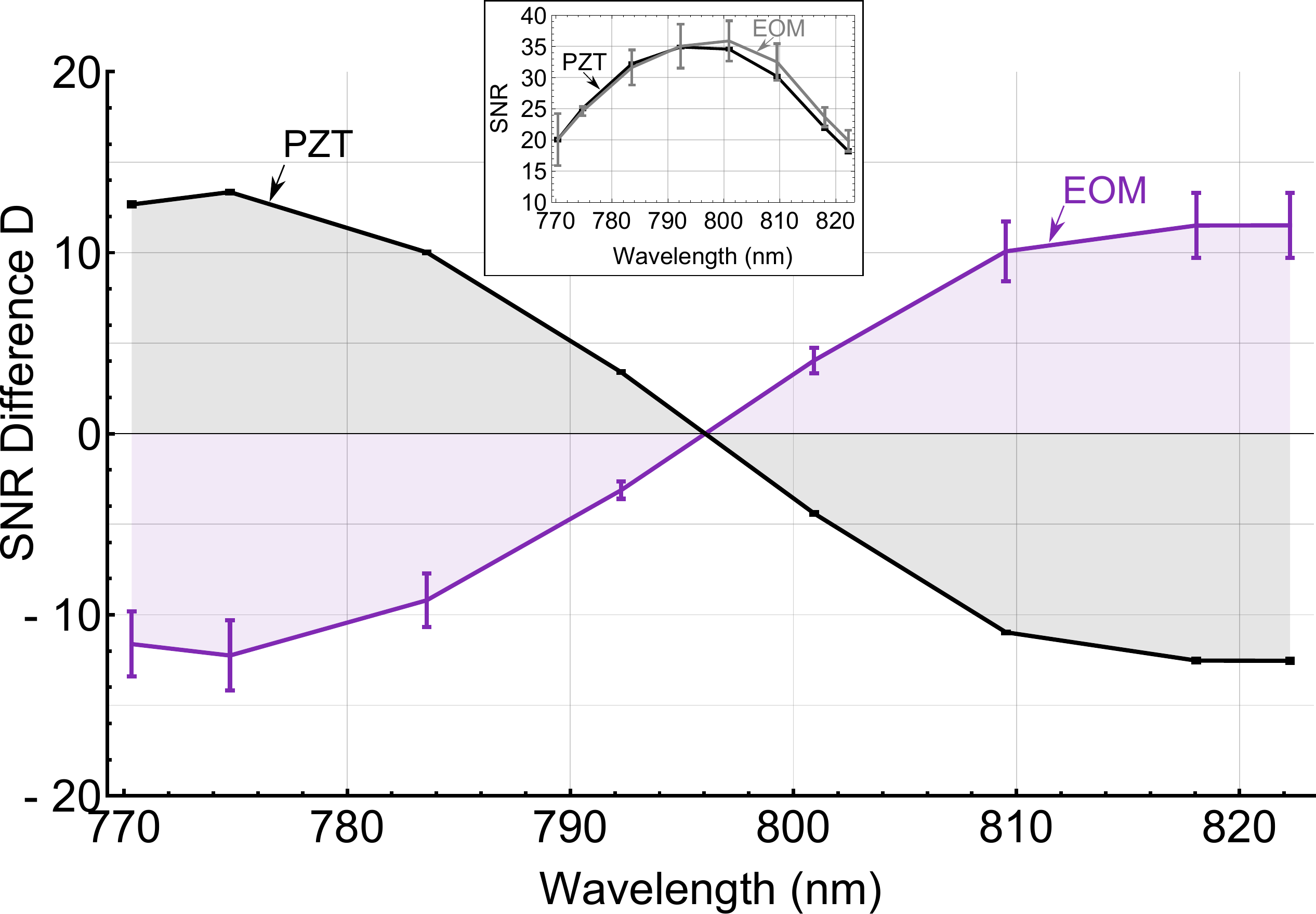}
	\caption{
		Signal-to-noise ratio for a phase modulation of an EOM and a piezo actuator (inset) where the dominant contribution of the phase mode has been removed. Error bars: standard deviation.
	}
	\label{fig:dispersion}
\end{figure}

Having established the principle of mode-specific measurements in a dispersion-free medium (i.e., air), it is interesting to consider the spectral form of the homodyne signal from a perturbation with a frequency-dependent modulation depth (i.e., a dispersive medium). In such a case, the temporal displacements of the carrier $t_\varphi$ and that of the envelope $t_g$ are no longer equivalent due to the material dispersion (i.e., $n'(\w_0)$) as seen in Eq.~\ref{eq:fieldexpanded2}. Hence, projection of the homodyne signal onto the carrier and group modes do not reveal the same, albeit scaled, information regarding the temporal / distance fluctuation. Rather, there is an additional temporal displacement of the envelope with respect to the carrier that arises from the $n'(\w_0)$ not present in the case of a modulation in air, and the ability to measure this relative temporal displacement can provide information about the medium in which the perturbation occurs. 

One candidate for a system in which the modulation depth is frequency dependent is an electro-optic modulator (EOM). Towards that aim, we performed an experiment in which an EOM is modulated in addition to the previously discussed piezo. The modulation frequency of the EOM is shifted from that of the piezo by $ \sim 1 \textrm{kHz}$, which allows both signals to be measured simultaneously. An additional digital demodulation at the $1\textrm{kHz}$ offset frequency for the heterodyned EOM signal is performed to separate the two homodyne signals. The piezo homodyne signal thus provides a reference to which the EOM signal is compared, and any deviation from the spectral form of the piezo signal is indicative of a supplemental dispersive effect. 

The retrieved SNRs for the EOM and piezo modulations are shown in the inset of Fig~\ref{fig:dispersion} where an overall scaling factor has been added to the piezo signal (the factor is $G_{\textrm{PZT}}$ as defined in the ensuing discussion). 
	As already discussed, each of these signals resembles the static mean signal field, which indicates that the carrier mode contribution is dominant. Following a procedure identical to that described in Section~\ref{sec:multimode}, the carrier mode contribution is subtracted from these two signals by using the DC spectrum as a reference for the mean field, i.e., $u_{0}(\W) \propto \sqrt{P_{s}(\W)}$.
	As such, the difference spectrum is constructed for the EOM signal according to $D_{\textrm{EOM}}(\W) = \Sigma_{\textrm{EOM}} - G_{\textrm{EOM}} \cdot \sqrt{P(\W)}$ where $G_{\textrm{EOM}} = \int d\W \, \Sigma_{\textrm{EOM}} \, \sqrt{P(\W)} / \int d\W \, P(\W)$. An analogous difference spectrum is computed for the piezo $D_{\textrm{PZT}}(\W)$, in which an additional scaling factor is included that serves to equate the carrier mode contributions from the two modulations 
	 \footnote{ $D_{\textrm{PZT}}(\W) = \frac{ G_{\textrm{EOM}} }{  G_{\textrm{PZT}} } \cdot \left[ \Sigma_{\textrm{PZT}} - G_{\textrm{PZT}} \cdot \sqrt{P(\W)} \right] $}. The difference signals for the two types of perturbations are shown in Fig~\ref{fig:dispersion}. 
	
	As seen in Fig~\ref{fig:dispersion}, the difference spectrum for the piezo element (black trace) is similar to what was observed in Fig.~\ref{fig:diffsignal} and resembles a time-of-flight mode $u_{g}(\W) = (\W / \Delta \w) \cdot u_{0} (\W)$. The difference spectrum for the EOM, however, is clearly of a different character and has approximately an equivalent magnitude but opposite sign. The difference between these two spectra indicates that the modulation depth for the EOM is indeed spectrally dependent. Hence, the group mode projection allows one to reliably differentiate a change in the absolute propagation distance from that of a dispersive variation of the medium. 
		
The difference between the two spectra of Fig~\ref{fig:dispersion} may be used to quantify the magnitude of dispersion observed with the EOM. The fact that the difference spectrum for the EOM largely mirrors that of the dispersion-free PZT indicates that the EOM modulation depth varies linearly across the bandwidth. The difference between these two spectra is written as: $ \Delta \phi _{\textrm{eom}}/ \Delta \omega = \left[ \left. d \phi_{\textrm{eom}}(\W) / d \w \right|_{\w_0} - \left. d \phi_{\textrm{pzt}}(\W) / d \w \right|_{\w_0} \right] = \left( \gamma - 1 \right) \cdot \left. d \phi_{\textrm{pzt}}(\W) / d \w \right|_{\w_0} $. The coefficient $\gamma$ is the proportionality constant between the two spectra of Fig~\ref{fig:dispersion} and is determined by minimizing the expression $ \min \left(   \int d \omega \left[ \gamma \cdot D_{\textrm{EOM}}(\W) - D_{\textrm{PZT}}(\W) \right]^{2}  \right) $. A proportionality constant of $\gamma = -0.82 \pm 0.15$ relates the two traces of Fig~\ref{fig:dispersion}, which agrees with the observation that the difference spectrum for the EOM is approximately of equal magnitude with that of the piezo. 
It is then straightforward to show that for a length modulation in a dispersion free medium, the linear coefficient of the spectral phase may be related to the zeroth-order spectral phase through the relation $ \left. d \phi_{\textrm{pzt}}(\W) / d \w \right|_{\w_{0}} = \phi_{0,\textrm{pzt}} / \w_{0}$ \footnote{ The spectral phase for a length perturbation in a dispersion free medium is given as $\phi(\w) = \left(\w_{0} / c \right) \delta L + \left(\W / c \right) \delta L$. Identifying the terms $ \phi_{0} = \left(\w_{0} / c \right) \delta L$ and $ \left. d \phi(\W) / d \w \right|_{\w_{0}}= \delta L / c$, it follows that $\left. d \phi(\W) / d \w \right|_{\w_{0}}= \phi_{0} / \w_{0}$ }.
Therefore, the variation of the modulation depth across the bandwidth relative to that of a dispersion-free modulation is finally written as $\Delta \phi_{\textrm{eom}} / \phi_{0,\textrm{pzt}}$, and is found to be: $\Delta \phi_{\textrm{eom}} / \phi_{0,\textrm{pzt}} \simeq \left( \gamma - 1 \right) \cdot 2 \sigma / \w_{0} $. Given a proportionality coefficient of $\gamma = -0.82 \pm 0.15$ between the difference spectra, the variation of the EOM's modulation depth across the optical bandwidth is $\sim ( 7 \pm 1 ) \%$.
Consequently, the group mode $u_{g}(\W)$ also allows one to quantify the change in the dispersive properties of a medium seen by the signal field. 
\section{Conclusion}

In summary, a modal description is presented of a multispectral laser field whose longitudinal propagation is disrupted. The analyzed light is spectrally dispersed onto a photodiode array, which allows for simultaneously observing the effect of the displacement on up to eight distinct frequency bands. Various linear combinations of these bands are formed, which reveal the temporal displacement experienced by both the optical carrier and the slowly-varying envelope. Forming specific linear combinations of these photocurrents is equivalent to projecting out a particular modal pattern from a larger multidimensional object. We have shown that these projective measurements are capable of retrieving displacement information down to the shot noise limit and exhibit a sensitivity that is in good agreement with theoretical predictions. 

	Moreover, by combining the information contained within a phase and time-of-flight measurement, it is possible to define an additional detection mode, which possesses a sensitivity higher than that afforded by only a standard interferometric detection. The precision of this hybrid detection mode is optimal, meaning that it corresponds to the ultimate detection sensitivity available with classical light resources. 
	
	The experimental strategies outlined in this work serve as a proof-of-concept for multimodal detection of femtosecond pulse trains. Importantly, these techniques are not limited to the measurement of longitudinal displacements, but may be applied for the retrieval of more complex parameters embedded within the field, including dispersive properties or even absorption. These techniques are also readily applicable to characterizing a source's intrinsic timing jitter and may be used to extract the relative jitter between two identical sources or a single source with the use of a fiber  delay. Moreover, it is feasible to utilize multispectral detection with non-classical, frequency entangled quantum resources for either quantum-enhanced metrology or the creation of quantum multimode states. Finally, the use of these techniques is not isolated to Ti:sapphire systems and silicon photodiodes. They may be adapted to alternative wavelength ranges through an appropriate choice of photodiode and spectrally-dispersive element. 
	
In conclusion, spectrally-resolved multimode detection offers an interesting measurement strategy that is readily-implementable, adaptable, sensitive, and capable of interrogating light fields on a variety of relevant timescales.

\bigskip

The work was supported by the European Research Council starting grant Frecquam. J.R. acknowledges support from the European Commission through Marie Curie Actions QOCO. N.T. and C.F. are members of the Institut Universitaire de France. This work has received funding from the European Union’s Horizon 2020 research and innovation program under Grant Agreement No. 665148.
\bibliographystyle{apsrev}
\bibliography{timing-paper-bib}

\end{document}